\documentclass[12pt]{article}
\usepackage{latexsym}
\usepackage{amsmath,amsbsy,amssymb}

\usepackage{curves}
\usepackage{epic}
\usepackage{eepic}
\usepackage{epsfig}

\newcommand{\dd}{\mbox{\rm d}}

\newcommand{\gam}{\gamma}

\newcommand{\tl}{\tilde}

\newcommand{\ddt}{\frac{\dd}{\dd \tau}}
\newcommand{\hmu}{{\hat \mu}}
\newcommand{\dotx}{{\dot x}}

\newcommand{\p}{\partial}

\newcommand{\be}{\begin{equation}}
\newcommand{\bear}{\begin{eqnarray}}
\newcommand{\ear}{\end{eqnarray}}
\newcommand{\ee}{\end{equation}}

\newcommand{\lbl}{\label}
\newcommand{\bi}{\bibitem}
\newcommand{\ci}{\cite}

\newcommand{\vs}{\vspace}
\newcommand{\hs}{\hspace}

\begin{document}

\begin{center}

\

\vs{.8cm}

\baselineskip .7cm

{\bf \Large A Non-Trivial Zero Length Limit\\
 of the Nambu-Goto String } 

\vs{4mm}

\baselineskip .5cm
Matej Pav\v si\v c

Jo\v zef Stefan Institute, Jamova 39,
1000 Ljubljana, Slovenia

e-mail: matej.pavsic@ijs.si

\vs{3mm}

{\bf Abstract}
\end{center}

\baselineskip .43cm
{\small
We show that a Nambu-Goto string has a nontrivial zero length limit which
corresponds to a massless particle with extrinsic curvature. The
system has the set of
six first class constraints, which restrict the phase space variables so that
the spin vanishes. Upon quantization, we obtain six conditions on the state,
which can be represented as a wave function of position coordinates, $x^\mu$,
and velocities, $q^\mu$. We have found a wave function $\psi(x,q)$ that turns
out to be a general solution of the
corresponding system of six differential equations, if the dimensionality
of spacetime is eight. Though classically the system is just a point
particle with vanishing extrinsic curvature and spin, the quantized system
is not trivial, because it is consistent in eight, but not in
arbitrary,  dimensions. 
}

\vs{3mm}

\hs{7mm}

\baselineskip .55cm

\section{Introduction}

Before the occurrence of string theories, elementary particles were
described as point particles. They could live in principle in arbitrary
dimensions, interacting by gravitational and other three fumdamental
forces. In string theories, particles and fields are excitations of
a string. Quantized bosonic string theory is consistent in 26-dimensions.
As an approximation, a string can be treated as a point particle with
extrinsic curvature and spin\,\ci{Lindstrom2,PavsicRigid1,PavsicRigidFromString},
the so called rigid particle\,\ci{Nesterenko1}--\ci{PavsicRigidRevisited}. 
In the description of Ref.\,\ci{PavsicRigidFromString}, the system has two first
class constraints, inherited from the string, and four additional constraints
that are second class. In this paper we consider the zero length limit
of such a system, in which case all six constraints become first class, and
effectively eliminate from the description all the degrees of freedom, except
those of a point particle, whose extrinsic curvature and spin vanish.
At first sight this could mean that we have arrived at the theory of a point
particle, living, in principle, in arbitrary dimensions. But the six
first class constraint are still present there, and upon quantization, they
become restrictions on possible physical states. We have found that for
a rather general class of solutions the quantum description can be
performed consistently in eight dimensions, but not in other dimensions.

In Sec.\,2 we derive a particle with extrinsic curvature from a string, and
in Sec.\,3 we consider its zero length limit. We obtain the same action that
had already been considered by McKeon\,\ci{McKeon}. However, in distinction
to the case of Ref.\,\ci{McKeon}, our system is subjected to a
constraint, inherited from the string theory, that was not taken into account
in Ref.\,\ci{McKeon}. Therefore, our dynamical system is different, because
it has two primary constraints, whose conservation gives additional
four constraints. Altogether, we obtain six constraints that turn out to be
all first class. In the presence of those constraints, the particle's center
of mass momentum $p^\mu$,
velocity $q^\mu={\dot x}^\mu$, and the conjugate momentum $\pi^\mu$ are
all parallel to each other. Therefore, the particle's spin and extrinsic
curvature are zero, which means that the particle's position $x^\mu (\tau)$
describes a straight worldline, and not a helix, as in the case of a
rigid particle. In Sec.\,4 we quantize the system by imposing the six
constraints as restrictions on physical states, and find a wave function
that solves the latter system of equations, provided that the dimension
of the space in which the particle lives, is eight. In Conclusion we
argue why this is a remarkable, nontrivial, result, revealing
yet another surprising property of string theories.

\section{The particle with curvature from a string}

In the previous paper\,\ci{PavsicRigidFromString} its was shown that
one can obtain a particle with curvature as an approximation to a string,
living in a target space with an extra time like dimension. The string
equation of motion in the conformal gauge are then
\be
  {\ddot X}^{\hat \mu} + X''^{\hat \mu} = 0,~~~~
  {\dot X}^{\hat \mu}{\dot X}_{\hat \mu} - X'^{\hat \mu} X'_{\hat \mu}=0,
  ~~~~{\dot X}^{\hat \mu} X'_{\hat \mu} = 0,
\lbl{2.1}
\ee
where ${\hat \mu} = (\mu,D+1)$, $\mu=0,1,2,3,...,D-1$. A possible
solution is
\bear
  &&X^{\mu} = C^\mu +\sum_n (a_n^\mu {\rm cos} \, \omega_n \tau
        + b_n^\mu {\rm sin} \, \omega_n \tau ) {\rm e}^{k_n \sigma} \nonumber \\
  &&X^{D+1} = \sigma~,~~~~~\sigma \in [0,L], \lbl{2.2}
\ear
where
\bear
  &&\omega_n^2-k_n^2=0,~~~a_n^2=b_n^2,~~~C_\mu a_n^\mu=C_\mu b_n^\mu =
  a_n^\mu b_{n \mu}=0,\nonumber\\
  &&C^2=1. \lbl{2.3}
\ear

In particular, if all higher modes with $n\ge 1$ vanish, we have:
\bear
  &&X^{\mu} = C^\mu +(a^\mu {\rm cos} \, \omega \tau
        + b^\mu {\rm sin} \, \omega \tau ) {\rm e}^{k \sigma} \nonumber \\
  &&X^{D+1} = \sigma~,~~~~~\sigma \in [0,L], \lbl{2.4}
\ear
where we have denoted $a_1^\mu \equiv a^\mu,~b_1^\mu \equiv b^\mu,~
\omega_1 \equiv \omega$. Such a string satisfies the Dirichlet boundary
condition
\be
   \delta X^{\hat \mu}|_B = 0,
\lbl{2.5}
\ee
such that the string ends move on a $D$-brane\,\ci{PavsicRigidFromString}.

For a fixed $\sigma$, Eq.\,(\ref{2.4}) describes a helix in $D$-dimensions.
If the string length $L$ is small in comparision with the radius of the
helix, then the string effectively behaves like a point-particle, tracing
a helical worldline.

The string embedding functions can be expanded according to\,\ci{Lindstrom2,
PavsicRigid1,PavsicRigidFromString},
\bear
  &&X^\mu (\tau,\sigma) = x^\mu (\tau) + y^\mu (\tau) k \sigma +
  {\cal O}(k^2 \sigma^2) \nonumber \\
  &&X^{D+1} (\sigma) = \sigma, \lbl{2.6}
\ear
where $k$ is a constant. For the solution (\ref{2.4}) this gives
\bear
  x^\mu (\tau) = C^\mu \tau + a^\mu {\rm cos}\, \omega \tau +
    b^\mu {\rm sin}\,\omega \tau \nonumber\\
  y^\mu = a^\mu {\rm cos}\, \omega \tau
   + b^\mu {\rm sin}\,\omega \tau . \lbl{2.7}
\ear

From now on, we will consider the expansion (\ref{2.6}), and search for
the action satisfied by the variables $x^\mu (\tau)$ and $y^\mu (\tau)$.
In Ref.\,\ci{PavsicRigidFromString} we started from the Polyakov action
\be
     I[X^\hmu,\gam^{ab}] = \frac{T}{2} \int \dd^2 \xi \, \sqrt{ \gam}
     \, \gam^{ab} \p_a X^\hmu \p_b X_\hmu ,
\lbl{2.8}
\ee
where $T$ is the string tension, and $\xi^a = (\tau,\sigma)$.

Using the expansion (\ref{2.6}), the action (\ref{2.8})
becomes\,\ci{PavsicRigidFromString}
\be
     I = \frac{LT}{2} \int \dd \tau \, \left [ \frac{1}{e} ({\dot x}^2 + L k
   {\dot x}{\dot y}) + e (1+f^2) (k^2 y^2 + 1) - 2 f k {\dot x} y \right ]
   + {\cal O}(k^2 L^2) ,
\lbl{2.9}
\ee
where $e(\tau)$ and $f(\tau$ comes from the expansion of $\sqrt{\gam}\gam^{11}$
and $\sqrt{\gam} \gam^{12}$, respectively, whereas the expansion of
$\sqrt{\gam} \gam^{22}$ gives $e(\tau) (1 + f^2(\tau)) + {\cal O}(\sigma)$.
The equations of motion are:
\bear
 &&\delta e\; : ~~~~- \frac{1}{e^2} ({\dot x}^2 + L k{\dot x}{\dot y}) +
         (1+f^2) (1+ k^2 y^2) = 0, \lbl{2.13a}\\
 &&\delta f\; : ~~~~ f e (1+k^2 y^2)- k {\dot x} y = 0, \lbl{2.14} \\
 && \delta y\; : - L k \, \frac{\dd}{\dd \tau}
  \left ( \frac{{\dot x}^\mu} {e} \right )
     + 2 e (1+f^2) y^\mu - 2 f k {\dot x}^\mu = 0,  \lbl{2.15}\\
 &&\delta x\; : ~~~~\frac{\dd}{\dd \tau} \left ( \frac{{\dot x}^\mu}{e}
   + \frac{L k {\dot y}^\mu}{2 e} - f k y^\mu \right ) = 0. \lbl{2.16}
\ear

In a gauge in which $f=0$, the action (\ref{2.9}) is
\be
   I = \frac{L T}{2} \int \dd \tau \, \left [ \frac{{\dot x}^2}{e}+e +
   \frac{Lk {\dot x}{\dot y}}{e} + e k^2 y^2 \right ].
\lbl{2.10}
\ee
If we plug the equation of motion
\be
   y^\mu = \frac{L}{2 k} \,\frac{1}{e}\, \frac{\dd}{\dd \tau}
           \left ( \frac{{\dot x}^\mu}{e} \right ),
\lbl{2.11}
\ee
and introduce the parameters
\be
  m=LT~,~~~~~\mu= \frac{L^3 T}{8},
\lbl{2.12}
\ee
then the action (\ref{2.10}) becomes\,\ci{PavsicRigidFromString}:
\be
   I[x^\mu,e] = \int \dd \tau \, \left [ \frac{m}{2} \left ( 
  \frac{{\dot x}^2}{e}+e \right ) - \frac{\mu}{e} \, \frac{\dd}{\dd \tau}
  \left (\frac{{\dot x}^\mu}{e} \right ) \frac{\dd}{\dd \tau}
  \left (\frac{{\dot x}_\mu}{e} \right ) \right ] .
\lbl{2.13}
\ee
This action contains first and second order derivatives of the variables
$x^\mu (\tau)$. According to the Ostrogradski formalism\,\ci{Ostrogradski},
higher derivative theories contain negative energies. In the presence
of an interaction such a system can become unstable. A relatively
recent finding is that this is not always so. As shown in
Refs.\,\ci{Smilga}--\ci{Ilhan}, there exist interacting second order systems
that are unconditionally stable. Moreover, as pointed out by
Woodard\,\ci{Woodard}, the presence of a sufficient number of gauge constraints
can stabilize the system. As an example, Woodard cited the second derivative
model of a a massless point particle with rigidity, considered by
Plyushchay\,\ci{Plyushchay1}.

\section{Zero length limit}

We will now consider the limit in which the string length $L$ tends to
zero. For such purpose let us introduce a new parameter $\tau' = h (\tau)$,
and a new Lagrange multiplier ${\tl e} (\tau')$ according to the
relation
\be
  \dd \tau e = \dd \tau' m {\tl e}.
\lbl{3.1}
\ee
Under such a reparametrization the action (\ref{2.13}) becomes
\be
   I[x^\mu,{\tl e}] = \int \dd \tau' \, \left \{ \frac{1}{2} \left [ 
  \left ( \frac{\dd x}{\dd \tau'} \right )^2 \frac{1}{\tl e}
  + m^2 {\tl e} \right ] - \frac{\mu}{m^3} \frac{1}{\tl e} \frac{\dd}{\dd \tau'}
  \left (\frac{1}{\tl e} \frac{\dd x^\mu}{\dd \tau'}  \right )
  \frac{\dd}{\dd \tau'}\left (\frac{1}{\tl e} 
  \frac{\dd x^\mu}{\dd \tau'}  \right ) \right \}.
\lbl{3.2}
\ee
The parameter $\tau'$ can be renamed into $\tau$, and the latter action can
be written as
\be
   \int \dd \tau \left [ \frac{1}{2} \left ( \frac{{\dot x}^2}{\tl e}
   + m^2 {\tl e} \right ) - \frac{\mu}{m^3 {\tl e}} \frac{\dd}{\dd \tau}
   \left ( \frac{{\dot x}^\mu}{\tl e} \right ) \frac{\dd}{\dd \tau}
   \left ( \frac{{\dot x}_\mu}{\tl e} \right ) \right ].
\lbl{3.3}
\ee
Expressing $m$ and $\mu$ according to Eq.\,(\ref{2.12}), the coefficient
in front of the second term of the latter action becomes
$\mu/m^3 = 1/(8 T^2) \equiv {\tl \mu}$. In Eq.\,(\ref{3.3}) we have
a term that corresponds to the Howe-Tucker action\,\ci{Howe-Tucker}, and
an extra term that corresponds to the particle's curvature.

The action (\ref{3.3}) has two important limits:

\ (i) $T \rightarrow \infty$, implying $\mu/m^3 \equiv {\tl \mu} = 1/(8T^2)
\rightarrow 0$. The term with curvature then disappears from the action. The
term containing $m = LT$ would become infinite, unless we also impose the limit
$L \rightarrow 0$ such that $m=LT$ remains finite. Then Eq.\,(\ref{3.3})
becomes the well-known Howe-Tucker action for massive point particle.

(ii) $T$ finite, $L \rightarrow 0$. In such limit, we have $m=LT \rightarrow 0$,
whereas $\mu/m^3 \equiv {\tl \mu} = 1/(8 T^2)$
remains intact, and the action (\ref{3.3}) becomes
\be
   I[x^\mu,{\tl e}] = \int \dd \tau \, \left [ \frac{1}{2 {\tl e}}
  {\dot x}^2 - \frac{{\tl \mu}}{\tl e} \, \frac{\dd}{\dd \tau}
  \left (\frac{{\dot x}^\mu}{\tl e} \right ) \frac{\dd}{\dd \tau}
  \left (\frac{{\dot x}_\mu}{\tl e} \right ) \right ] .
\lbl{3.4}
\ee
The latter action is identical to the action for the ``massless" particle
with curvature, considered by McKeon\,\ci{McKeon}.

In the following we will investigate in some detail the case (ii).
From now on, we will rename ${\tl e}$ into $e$, and ${\tl \mu}$ into $\mu$,
and write the action (\ref{3.4}) as
\be
   I[x^\mu,{e}] = \int \dd \tau \, \left [ \frac{1}{2 {e}}
  {\dot x}^2 - \frac{{\mu}}{e} \, \frac{\dd}{\dd \tau}
  \left (\frac{{\dot x}^\mu}{e} \right ) \frac{\dd}{\dd \tau}
  \left (\frac{{\dot x}_\mu}{e} \right ) \right ] .
\lbl{3.5}
\ee

The canonical momenta are
\bear
  &&p_\mu = \frac{\p L}{\p {\dot x}^\mu} 
  - \ddt \left (\frac{\p L}{\p {\ddot x}^\mu} \right )
  = \frac{{\dot x}_\mu}{e} + \frac{2 \mu}{e}\, \ddt \left ( \frac{1}{e}\,
  \ddt \left ( \frac{{\dot x}_\mu}{e} \right ) \right ) , \lbl{3.6} \\
  &&\pi_\mu = \frac{\p L}{\p {\ddot x}^\mu} = - \frac{2 \mu}{e^2} \,
  \ddt \left ( \frac{{\dot x}_\mu}{e} \right ) , \lbl{3.7} \\
  &&p_e = \frac{\p L}{{\p \dot e}} = \frac{2 \mu}{e^3}\, {\dot x}^\mu \, \ddt
  \left ( \frac{{\dot x}_\mu}{e} \right ) .   \lbl{3.8}
\ear

The equations of motion are
\bear
  &&\delta x^\mu\,: ~~~~{\dot p}_\mu = 0 \lbl{3.9}\\
  &&\delta e\,: ~~~~\frac{\p L}{\p e} - \ddt \frac{\p L}{\p{\dot e}} \nonumber\\
  &&~~~~~~~\frac{{\dot x}^2}{e^2}  + 3 \mu \,\frac{1}{e}
   \ddt \left ( \frac{{\dot x}^\mu}{e} \right )\frac{1}{e}
   \ddt \left ( \frac{{\dot x}_\mu}{e} \right ) - \frac{2 \mu}{e}\,
   \ddt \left ( \frac{{\dot x}^\mu}{e^2} 
   \ddt \left ( \frac{{\dot x}_\mu}{e} \right )  \right ) = 0. \lbl{3.10}
\ear
The Hamiltonian is
\be
   H_0 = p_\mu \dotx^\mu + \pi_\mu {\ddot x}^\mu + p_e {\dot e} - L_0 ,
\lbl{3.11}
\ee

Let us introduce the new variables
\be
   \dotx^\mu = q^\mu,~~~~~~{\dot e} = \beta .
\lbl{3.12}
\ee
From Eqs.\,(\ref{3.6}),(\ref{3.7}) we have
\be
   {\ddot x}^\mu = \frac{e^3}{2 \mu} \pi^\mu + \frac{{\dot e}}{e} q^\mu,
   ~~~~~ p_e = - \frac{\pi_\mu q^\mu}{e},
\lbl{3.13}
\ee
and after inserting the latter expressions into the Hamiltonian (\ref{3.11}),
we obtain
\be
  H_0 = e \left ( \frac{p_\mu q^\mu}{e}- \frac{e^2 \pi^2}{4 \mu}
  -\frac{q^2}{2 e^2}\right ) + \beta ( p_e + \frac{\pi_\mu q^\mu}{e} ).
\lbl{3.14}
\ee

In deriving the action (\ref{2.13}) we used  a gauge in which $f=0$.
In such a gauge the constraint (\ref{2.14}) becomes
\be
  {\dot x}^\mu y_\mu = 0.
\lbl{3.15}
\ee
By using Eqs.\,(\ref{2.11}),(\ref{3.7}) and (\ref{3.12}), the latter
equation can be written as
\be
  \pi_\mu q^\mu = 0.
\lbl{3.16}
\ee
Our action (\ref{2.13}) and its $L \rightarrow 0$ limit (\ref{3.5}) is then
subjected to the constraint (\ref{3.16}). Therefore, the Lagrangian
${\cal L}_0$ must be supplemented with the above constraint:
\be
  {\cal L} = {\cal L}_0 - \alpha \pi_\mu q^\mu ,
\lbl{3.17}
\ee
and the Hamiltonian $H_0$ with
\be
  H = H_0 + \alpha \pi_\mu q^\mu.
\lbl{3.18}
\ee

The equations of motion derived from the Hamiltonian $H$ are
\bear
  &&\dotx^\mu =\lbrace x^\mu, H \rbrace = q^\mu, \lbl{3.19}\\
  &&{\dot e} = \lbrace e, H \rbrace = \beta , \lbl{3.20}\\
  &&{\dot q}^\mu = \lbrace q^\mu,H \rbrace = - \frac{e^3 \pi^\mu}{2 \mu}
  + \alpha q^\mu + \frac{\beta q^\mu}{e} , \lbl{3.21}\\
  &&{\dot p}_\mu = \lbrace p_\mu , H \rbrace = 0 ,  \lbl{3.22}\\
  &&{\dot \pi}_\mu = \lbrace \pi_\mu, H \rbrace =
   - \left ( p_\mu - \frac{q_\mu}{e} + \alpha \pi_\mu
    + \frac{\beta \pi_\mu}{e} \right ) , \lbl{3.23}\\
  &&{\dot p}_e = \lbrace p_e , H \rbrace = - \frac{3 e^2 \pi^2}{4 \mu}
  +\frac{q^2}{2 e^2} - \beta \frac{\pi_\mu q^\mu}{e^2}. \lbl{3.24}
\ear

Variation of the action $\int {\cal L} \dd \tau$ with respect to
$e$ and $\alpha$, gives the constraints
\bear
  &&{\tl \phi}_1 = \frac{3 e^2 \pi^2}{4 \mu} 
   - \frac{q^2}{2 e^2} = 0 , \lbl{3.25}\\
  &&\phi_2 = \pi_\mu q^\mu = 0 , \lbl{3.26}
 \ear

From the requirement that those constraints
must be preserved in time, we obtain another three constraints,
\bear
  && \phi_3 = e p_\mu \pi^\mu, \lbl{3.28}\\
  &&\phi_4 = \frac{p_\mu q^\mu}{e}
    +\frac{e^2 \pi^2}{2 \mu}- \frac{q^2}{e^2}, \lbl{3.29}\\
   &&\phi_5 = p^2 - \frac{p_\mu q^\mu}{e} \lbl{3.30}.
\ear

The linear combination
\be
  {\phi}_1 = - {\tl \phi}_1 + \phi_4 = \frac{p_\mu q^\mu}{e} -
   \frac{q^2}{2 e^2}  - \frac{e^2 \pi^2}{4 \mu} 
\lbl{3.31}
\ee
is an expression that enters the Hamiltonian (\ref{3.14}).

Variation of the action $\int {\cal L} \dd \tau$ with respect  $\beta$
gives the constraint
\be
  \phi_6 = e p_e + \pi_\mu q^\mu = 0 . \lbl{3.27}
\ee
The equation ${\dot \phi_6}=\{\phi_6,H\}$ for conservation of $\phi_6$
does not give a new constraint.

The constraints $\phi_i$. $i=1,2,3,4,5$, include only the momenta $p_\mu$ and
$\pi_\mu$, conjugated to the dynamical variables $x^\mu$ and $q^\mu$,
whereas the constraint $\phi_6$ includes also the momentum $p_e$ which,
due to (\ref{3.26}) vanishes on the constraint surface. Thus, $p_e$ is
merely an auxiliary momentum, and the constraint associated with it is treated
as the ``last" one.

From $\phi_1 =0$, $\phi_4 =0$, $\phi_5 =0$ it follows that
\bear
    && \frac{p_\mu q^\mu}{e} = p^2, \lbl{3.32}\\
    && \frac{q^2}{e^2} = \frac{3}{2} p^2,
    \lbl{3.33}\\
    && \frac{e^2 \pi^2}{\mu} = p^2. \lbl{3.34}
\ear

Because $q^\mu$ is supposed to be a time like vector and $\pi^\mu$ a space
like vector,
and because their scalar product, $\pi_\mu q^\mu$, vanishes on the constraint
surface, it follows that $q^2 =0$, $\pi^2 \le 0$. Taking also into account that
Eq.\,(\ref{3.25}) implies the proportionality between $q^2$ and $\pi^2$,
it follows that $\pi^2 =0$. Eqs.\,(\ref{3.32})--(\ref{3.34} then become
\bear
  &&\frac{p_\mu q^\mu}{e} = 0, \lbl{3.35}\\
  &&\frac{q^2}{e^2} = 0,~~~~ \frac{e^2 \pi^2}{\mu} = 0, \lbl{3.36}
\ear
implying $p^2 =0$.

Because $p^2 = 0$, it follows that all constraints $\phi_i$, $i=1,2,...,6$,
are {\it first class}, i.e., $\lbrace \phi_i,\phi_j \rbrace = 0$. This can
be verified by calculating the Poisson brackets between all the constraints.
In fact, the constraints become $\phi_1=p_\mu q^\mu/e$, $\phi_2=q^\mu \pi_\mu$,
$\phi_3 = e p_\mu \pi^\mu$, $\phi'_4 = \pi^2$, $\phi'_5 = p^2$, $\phi'_6 = e p_e$,
where $\phi'_4$, $\phi'_5$ and $\phi'_6$ are the appropriate linear combinations
of the constraints $\phi_i$.

Eqs.\,(\ref{3.26}),(\ref{3.28}),(\ref{3.32}) and (\ref{3.34}) imply that
$q^\mu$, $\pi^\mu$ and $p^\mu$ are parallel. Consequently, the spin tensor
$S^{\mu \nu}=q^\mu \pi^\nu - q^\nu \pi^\mu$ vanishes. The parallelism between
$q^\mu = {\dot x}^\mu$ and $p^\mu$ means that the 4-velocity oscillations are
tangential to the worldline of the particle's center of mass. Therefore, the
center of mass worldline
\be
   x_{\rm T}(\tau) = x_0^\mu + p^\mu \tau ,
\lbl{3.37}
\ee
and the particle's position worldline
\be
  x^\mu (\tau) = x_0^\mu + p^\mu \tau + a^\mu {\rm cos}\, \omega \tau +
  b^\mu {\rm sin}\, \omega \tau ,
\lbl{3.38}
\ee
which are both solutions of the equations of motion (\ref{3.19})--(\ref{3.23}),
are not different wordllines. Both equations, (\ref{3.37}) and (\ref{3.38}),
represent the same curve, they differ only in the choice of parameter.
If in Eq.\,(\ref{3.37}) we change $\tau$ according to $\tau \rightarrow
1 + \alpha {\rm cos}\, \omega \tau + \beta {\rm sin}\, \omega \tau$, where
$\alpha$ and $\beta$ are proportionality factors, defined according to
$p^\mu = \alpha a^\mu$ and $p^\mu = \beta b^\mu$, we obtain
Eq.\,(\ref{3.38}).

The six first class constraints diminish the number of independent
degrees of freedom of our dynamical system. It turns out that $q^\mu$ and
$\pi^\mu$ are not dynamical degrees of freedom at all. Since $q^\mu$ and
$\pi^\mu$ are parallel to $p^\mu$, they bring nothing new to the classical
dynamics system. In the following we will investigate what happens if we
nevertleless pursue with the quantization of our constraint system.

\section{Quantization}

Upon quantization the phase space variables become the operators, satisfying
the commutation relations
\bear
&&[{\hat x}^\mu,{\hat p}_\nu] = i {\delta^\mu}_\nu,
~~~~[{\hat q}^\mu,{\hat \pi}_\nu] = i {\delta^\mu}_\nu, \lbl{4.1a}\\
&&[{\hat x}^\mu,{\hat x}^\nu]=0, ~~~[{\hat p}^\mu,{\hat p}^\nu]=0,
~~~[{\hat q}^\mu,{\hat q}^\nu]=0,~~~[{\hat \pi}^\mu, {\hat \pi}^\nu]=0,
\lbl{4.1b}
\ear 
and the constraints become restrictions on physical states:
\bear
  &&{\hat p}^\mu \frac{{\hat q}^\mu}{e}|\psi\langle =0, \lbl{4.2}\\
  &&\frac{1}{2}({\hat q}^\mu {\hat \pi}_\mu + {\hat \pi}_\mu {\hat q}^\mu)|\psi \rangle = 0, \lbl{4.3}\\
  &&e {\hat \pi}_\mu p^\mu |\psi \rangle = 0, \lbl{4.4}\\
  &&{\hat \pi}^\mu {\hat \pi}_\mu |\psi \rangle = 0, \lbl{4.5}\\
  &&{\hat p}^\mu {\hat p}_\mu |\psi \rangle = 0, \lbl{4.6}\\
  &&e {\hat p}_{\hat e}|\psi \rangle = 0. \lbl{4.7}
\ear
We do not impose the condition
\be
   {\hat q}^2 |\psi \rangle = 0,
\lbl{4.7a}
\ee
but only
\be
  \langle \psi |{\hat q}^2 |\psi \rangle = 0.
\lbl{4.7b}
\ee

In the representation in which ${\hat x}^\mu$ and ${\hat q}^\mu$ are diagonal, whereas
${\hat p}_\mu = - i \p/\p x^\mu$, ${\hat \pi}_\mu = -i \p/\p q^\mu$, Eqs.\,(\ref{4.6}) and
(\ref{4.5}) become massless Klein-Gordon equations in the $x^\mu$-space, and the
$q^\mu$-space, respectively.

A particular solution of Eqs.\,(\ref{4.6}),(\ref{4.5}) is
\be
  \psi_{p,q} (x^\mu,q^\mu) = {\rm e}^{i p_\mu x^\mu} {\rm e}^{i \pi_\mu q^\mu}
\lbl{4.8}
\ee
Here $p_\mu$ and $\pi_\mu$ are now eigenvalues of the corresponding
operators. The eigenvalues must satisfy the relations $p^\mu p_\mu = 0$ and
$\pi^\mu \pi_\mu=0$.

We will now show that a general solution of the system of equations
(\ref{4.2})--(\ref{4.7}) that satisfies the condition (\ref{4.7b}), is
\be
  \psi(x^\mu,q^\mu) = \int \dd^D p\, \dd^D \pi \, a(p,\pi) {\rm e}^{i p_\mu x^\mu}
  {\rm e}^{i \pi_\mu q^\mu} \delta(p^2) \delta (\pi^2) \delta (q^2)
  \delta(q^\mu \pi_\mu) \delta (p^\mu \pi_\mu) \delta (p_\mu q^\mu)
\lbl{4.9}
\ee  
where the constraints and the condition (\ref{4.7b}) are expressed in terms
of the $\delta$-functions.

\ \ \ (i) Eq.\,(\ref{4.2}) gives
\be
  {\hat p}_\mu {\hat q}^\mu \psi = \int \dd^D p\, \dd^D \pi a(p,\pi)
  p_\mu q^\mu \, {\rm e}^{i p_\mu x^\mu}
  {\rm e}^{i \pi_\mu q^\mu} \delta(p^2) \delta (\pi^2) \delta (q^2)
  \delta(q \pi) \delta (p \pi) \delta (p q) = 0,
\lbl{4.19}
\ee
because the integral of $p_\mu q^\mu \,\delta (p_\mu q^\mu)$ over $\dd^D p$
gives zero. 
We distinguish the operators from their eigenvalues by the hat symbol.

\ \ (ii) For Eq.\,(\ref{4.3}) we obtains
\bear
  &&({\hat q}^\mu {\hat \pi}_\mu - \frac{i}{2} D) \psi = 0 \lbl{4.20}\\
  &&{\hat q}^\mu {\hat \pi}_\mu \psi = (-i) \int \dd^D p \,\dd^D \pi a(p,\pi)
  {\rm e}^{i p_\nu x^\nu}\nonumber\\
  &&\hs{3cm} \times \frac{\p}{\p q^\mu} \Bigl(
  {\rm e}^{i \pi_\nu q^\nu} \delta(\pi^2) \delta (q \pi) \delta (q^2) 
  \delta(p q) \Bigr)
  \delta (p^2) \delta(p \pi).
\lbl{4.21}
\ear 
In Eq.\,(\ref{4.20}) we took into account the commutation relation
(\ref{4.1a}), which gives ${\hat \pi}_\mu {\hat q}^\mu =
{\hat q}^\mu {\hat \pi}_\mu - i D$.

We will use
\be
  \frac{\p}{\p q^\mu} \delta (f(q)) = \frac{\p f(q)}{\p q^\mu}
  \frac{\p \delta(f(q)}{\p f(q)},
\lbl{4.22}
\ee
which in particular gives
\be
  \frac{\p}{\p q^\mu} \delta(q^\nu \pi_\nu) 
  = \pi_\mu \frac{\p \delta(q^\nu \pi^\nu)}{\p (q^\nu \pi_\nu)}.
\lbl{4.23}
\ee
We then have
\bear
   &&\frac{\p}{\p q^\mu}\left (
  {\rm e}^{i \pi_\nu q^\nu} \delta (q \pi) \delta (q^2) \delta(p q)\right )
  = i \pi_\mu {\rm e}^{i \pi_\nu q^\nu} \delta (q \pi) \delta (q^2) \delta(p q)
  + \pi_\mu \frac{\delta(q \pi)}{\p (q \pi)} \delta(q^2) \delta(p q)
  \nonumber\\
  &&\hs{4cm} +
  2 q_\mu \frac{\p \delta(q^2)}{\p q^2} \delta(q \pi) \delta (p q)
   +
  p_\mu \frac{\p \delta(p q)}{\p (p q)} \delta (q \pi) \delta (q^2).
\lbl{4.24}
\ear
Inserting the latter expression into Eq.\,(\ref{4.21}), we obtain
\bear
  &&{\hat q}^\mu {\hat \pi}_\mu = (-i) \int \dd^D p \,\dd^D \pi \,a(p,\pi)
  {\rm e}^{i p_\nu x^\nu} {\rm e}^{i \pi_\nu q^\nu}\biggl(
  i q^\mu \pi_\mu  \delta (q \pi) \delta (q^2) \delta(p q)
  \nonumber \\
  &&\hs{3cm} + q^\mu \pi_\mu \frac{\delta(q \pi}{\p (q \pi)} \delta(q^2) \delta(p q) +
  2 q^\mu q_\mu \frac{\p \delta(q^2)}{\p q^2} \delta(q \pi) \delta (p q)
  \nonumber\\
  &&\hs{4cm} +
  q^\mu p_\mu \frac{\p \delta(p q)}{\p (p q)} \delta (q \pi) \delta (q^2)
  \biggr) \delta (p^2) \delta(p \pi) .
\lbl{4.25a}
\ear
Using the relation
\be
  x \delta'(x) = - \delta(x)
\lbl{4.25b}
\ee
we obtain
\be
  {\hat q}^\mu {\hat \pi}_\mu \psi = 4 i \psi.
\lbl{4.25}
\ee
Eq.\,(\ref{4.20}) then becomes
\be
   \left ( 4 i - \frac{i D}{2} \right ) \psi = 0,
\lbl{4.26}
\ee
which is satisfied if $D=8$.

\ (iii) Eq.\,(\ref{4.4}) gives:
\be
  {\hat p}^\mu \hat{\pi}_\mu \psi= \int \dd^D p \,\dd^D \pi \,a(p,\pi) p^\mu \pi_\mu
{\rm e}^{i p_\nu x^\nu}{\rm e}^{i \pi_\nu q^\nu} 
\delta(p^2) \delta (\pi^2) \delta (q^2)
  \delta(q^\mu \pi_\mu) \delta (p^\mu \pi_\mu) \delta (p_\mu q^\mu),
\lbl{4.27}
\ee
which vanishes, because of the expression $p^\mu \pi_\mu \delta (p^\mu \pi_\mu)$
under the integral.

\ (iv) In order to calculate Eq.\,(\ref{4.5}), we will use Eq.\,(\ref{4.24}),
in which we express the derivative of the $\delta$-function as
\be
  \delta' (x) = - \frac{\delta(x)}{x} + \Delta (x).
\lbl{4.27a}
\ee
The latter expression gives
\bear
  &&\int \dd x \, F(x) \delta' (x) = \int \dd x 
  \left ( F(0) + F'(x)\Bigl\vert_{x=0}\, x \right )
  \left ( - \frac{\delta(x)}{x} + \Delta (x) \right ) \nonumber\\
  &&\hs{2cm} = - F'(x)\Bigl\vert_{x=0} - \frac{F(x)}{x}\Bigl\vert_{x=0}+
  \int \dd x \, F(x) \Delta(x) = - F'(x)\Bigl\vert_{x=0},
\lbl{27b}
\ear
if we define $\Delta(x)$ according to
\be
  \int \dd x \, F(x) \Delta(x)=\frac{F(x)}{x}\Bigl\vert_{x=0},
\lbl{27c}
\ee
so that after the integration the term containing $\Delta(x)$ cancels out.
Then Eq.\,(\ref{4.24}) becomes
\bear
   &&\frac{\p}{\p q^\mu}\left (
  {\rm e}^{i \pi_\mu q^\mu} \delta (q \pi) \delta (q^2) \delta(p q)\right )
  = {\rm e}^{i \pi_\mu q^\mu}\delta (q \pi) \delta (q^2) \delta(p q)
  \left (2 i \pi_\mu - \frac{\pi_\mu}{q^\nu \pi_\nu} - \frac{2 q_\mu}{q^2}
  - \frac{p_\mu}{p^\nu q_\nu} \right )\nonumber \\
  &&\hs{7cm}+ ~terms~ with~ \Delta.
\lbl{27d}
\ear
If we use the above expression in Eq.\,(\ref{4.21}), we also obtain the
same result (\ref{4.25})

By using Eq.\,(\ref{27d}) in Eq.\,(\ref{4.5}), we obtain
\bear
 && {\hat \pi}^\mu {\hat \pi}_\mu \psi = - \frac{\p^2}{\p q^\mu \p q_\mu}
 = - \int \dd^D p\, \dd^d \pi \, {\rm e}^{i p_\nu x^\nu} 
 {\rm e}^{i \pi_\nu q^\nu} \delta (q \pi) \delta (q^2) \delta(p q) \nonumber\\
 &&\hs{3cm} \times \biggl\{ \left [ \pi_\mu \left (2 i - \frac{1}{q \pi} \right ) 
 - \frac{2 q_\mu}{q^2} - \frac{p_\mu}{p^\nu q_\nu} \right ]
\left [ \pi^\mu \left (2 i - \frac{1}{q \pi} \right )
 - \frac{2 q^\mu}{q^2} - \frac{p^\mu}{p^\nu q_\nu} \right ] \nonumber \\
 &&\hs{2cm} + \frac{\p}{\p q^\mu} \left ( - \frac{\pi^\mu}{q^2} - \frac{p^\mu}{p q}
 \right ) \biggr\} \nonumber \\
 &&\hs{1.5cm} = \frac{2}{q^2} (D-8) \psi .
\lbl{27e}
\ear
All other terms, including those with $\Delta$, vanish.

We have found that the constraint (\ref{4.5}) is satisfied by the wave
function (\ref{4.9}) in eight dimensions, just like the constraint
(\ref{4.3}).

\  (v) Eq.\,(\ref{4.6}) becomes
\be
  {\hat p}^\mu {\hat p}_\mu \psi = - \frac{\p^2 \psi}{\p x^\mu \p x_\mu}=0,
\lbl{4.29}
\ee
which vanishes because of the expression $p^2 \delta (p^2)$ under the
integral over $\dd^D p$.

\ \ (vi) Eq.\,(\ref{4.7}) becomes
\be
  - i \frac{\p}{\p e} \psi = 0,
\lbl{4.30}
\ee
which is fulfilled, because $\psi$ does not explicitly depend on $e$.

A remarkable feature of the above calculations is that the wave function
(\ref{4.9}) does not solve the quantum constraints (\ref{4.2})--(\ref{4.7})
and the condition (\ref{4.7b})
in arbitrary dimension $D$, but only in $D=8$. If (\ref{4.9}) is indeed the most general
solution of the system of equation (\ref{4.2})--(\ref{4.7}),(\ref{4.7b}), 
and there is no other solution, then
the system, obtained by
quantizing the zero length limit of the string, is consistent
in eight dimensions. Though the zero length limit is just like a point
particle, the system inherits from the string a set of constraints, which
upon quantization can be satisfied in eight dimensions, but not in an
arbitrary number of dimensions.

If we act on $\psi$ with the operator ${\hat S}_{\mu \nu} 
= {\hat q}_\mu {\hat \pi}_\nu - {\hat q}_\nu {\hat \pi}_\mu$, which is
the generator of rotations in the $q^\mu$-space, we obtain
\bear
   &&{\hat S}_{\mu \nu} \psi = \int \dd^D p \,\dd^D \pi \,
  \left [ (q_\mu \pi_\nu - q_\nu \pi_\mu ) \left ( 2 i - \frac{1}{q \pi}
  \right ) + \frac{1}{q^2} (q_\mu p_\nu - q_\nu p_\mu) \right ] \nonumber \\
  &&\hs{3.5cm} \times \,{\rm e}^{p_\rho x^\rho} {\rm e}^{\pi_\rho q^\rho}
   \delta(p^2) \delta (\pi^2)
  \delta(q \pi) \delta (q^2) \delta (p \pi) \delta (p q) 
\lbl{4.31}
\ear
The latter expression vanishes, because the $\delta$-functions restrict the
range of the variables $p^\mu$, $q^\mu$, $\pi^\mu$ on the surface, on which they
are all paralles to each other, so that on the surface, $q_\mu \pi_\nu - q_\nu \pi_\mu =0$
and $q_\mu p_\nu - q_\nu p_\mu = 0$.
The wave function $\psi(x^\mu,q^\mu)$ is thus a scalar under rotations
generated by ${\hat S}^{\mu \nu}$. The particle has vanishing spin.

\section{Conclusion}

We have found yet another surprising property of strings. So far it was
well known that a bosonic string can be consistently quantized in 26 dimensions,
but not in other dimensions\footnote{However, see Refs.\,\ci{PavsicSaasFee},
where slightly more general strings were shown to be consistent
in arbitrary dimensions.}. In this paper we considered a zero length
limit of a bosonic string. At first sight one would expect that such
a system is just a point particle, whose quantized counterpart can live
in arbitrary dimensions. But a thorough treatment of the constraints reveals,
that upon quantization we obtain a system of equations that can be solved
by a certain rather general wave function only in eight dimension. This means
that a quantized point particle that is obtained as a limit of a string must
live in eight dimensions, it cannot live in four dimensions. A consequence
is that, according to Kaluza-Klein theory, such a particle, in the case
when the 8-dimensional space is curved,
experiences the force that from the point of view of 4-dimensional
subspace manifests itself as gravitation and Yang-Mills forces. This means
that the zero point limit of the string leads to a theory that besides
gravitation contains other fundamental forces as well. 
The original string theory (of strings with finite extension)
also leads to gravitation and Yang-Mills fields,
though within a rather different theoretical procedure.

Zero length limit of a string and the corresponding theoretical description,
is merely a theoretical idealisation. In reality, a  string remains finite,
approximately being described as a zero length string living in eight
dimensions. In the approximate theory, only eight dimensions are
necessary for the consistency, the remaining eighteen dimensions are
superfluous. In fact the approximate theory is not consistent in
26-dimensions. The remaining eighteen dimensions are necessary for
consistent description of the remaining degrees of freedom that are
truncated in the approximate theory. Thus, treating a string as a point
particle, decouples eighteen dimenions from the description. The point particle
``sees" only eight dimensions, and, if the space is curved, feels gravitional
and Yang-Mills forces. Effectively, by treating the string approximately
as a point particle, we have reduced spacetime from twenty six to
eight dimensions, without really compactifying the remaining eighteen
dimensions; we have only eliminatied them from the dynamics, and thus
rendered them invisible to the particle. In other words, although
there might be present additional dimensions, the particle moves only
in an eihgt dimensional subspace.

\vs{4mm}

\centerline{\bf Acknowledgment}

This work has been supported by the Slovenian Research Agency.


\begin{thebibliography}{12}


\bi{Lindstrom2} J. Grundberg, J. Isberg, U. Lindstr\"om and H. Nordstr\"om,
Phys. Lett. B { 231} (1989) 61;
U. Lindstr\"om, Phys. Lett. B { 218} (1989) 315.

\bi{PavsicRigid1}
M. Pav\v si\v c,
  Class.\ Quant.\ Grav.\  {7} (1990) L187.
  
\bi{PavsicRigidFromString}
M. Pav\v si\v c,
``Point Particle with Extrinsic Curvature as an Approximation to a 
Nambu-Goto String: Classical and Quantum Model,''
  arXiv:1405.7838 [hep-th].
  
\bi{Nesterenko1} V.V. Nesterenko, J. Phys A { 22} (1989) 1673;\\
V.V. Nesterenko, Int. J. Mod. Phys. A { 6} (1991) 3989;\\
V. V. Nesterenko, S. H. Nguyen, Int. J. Mod. Phys. A {3} (1988) 2315;\\
B. P. Kosyakov, V. V. Nesterenko, Phys. Lett. B {384} (1996) 70.

\bi{Lindstrom3} J. Isberg, U. Lindstrom, H. Nordstrom and J. Grundberg,
Mod. Phys. Lett. A { 5} (1990) 2491.

\bi{Plyushchay} M. S. Plyushchay, Mod. Phys. Lett. A { 3} (1988) 1299;\\
Phys. Lett. B { 243} (1990) 383;\\
S. Klishevich, M.S. Plyushchay, Phys.Lett. B { 459} (1999) 201.

\bi{Plyushchay1}
M.S. Plyushchay,
Mod. Phys. Lett. A 4 (1989) 837.

\bi{Arodz} H. Arodz, A. Sitarzand and P. Wegrzin, Acta. Phys. Pol. B
{ 20} (1989) 921.

\bi{Nersessian} A. Nersessian, Theor. Math. Phys. { 117} (1998) 1214.

\bi{Ramos} E. Ramos, J. Roca, Nucl. Phys. B {436} (1995) 529.

\bi{Banerjee} Rabin Banerjee, Biswajit Paul, Sudhaker Upadhyay,
JHEP {08} (2011) 085.

\bi{Dereli} T. Dereli, D.H. Hartley, M. \"Onder and R.W. Tucker,
Phys. Lett. B {252} (1990) 601.

\bi{McKeon} D.G.C. McKeon, Class. Quant. Grav. {9} (1992) 2361.

\bi{Pavsic2}  M. Pav\v si\v c, Phys. Lett. B {221} (1989) 264.

\bi{Pavsic2a}
M. Pav\v si\v c,
  Class.\ Quant.\ Grav.\  {7} (1990) L187.
  
\bi{Nesterenko2} V.V. Nesterenko, A. Feoli and G. Scarpetta, J. Math. Phys.
{36} (1995) 5552.

\bi{PavsicRigidRevisited}
M. Pav\v si\v c,
  Found.\ Phys.\  {37} (2007) 40
  [hep-th/0412324].

\bi{Ostrogradski} M.V. Ostrogradski, Mem. Acad. Imper. Sci. St. Petersbg., 
{6} (1850) 385 .

\bi{Smilga} A.~V.~Smilga,
  SIGMA { 5} (2009) 017 
  [arXiv:0808.0139 [quant-ph]].

\bi{PavsicPU1}
M. Pav\v si\v c,
Mod.\ Phys.\ Lett.\ A {28} (2013) 1350165
  [arXiv:1302.5257 [gr-qc]].

\bi{PavsicPU2}
M. Pav\v si\v c,
  Phys.\ Rev.\ D {87} (2013) 10,  107502
  [arXiv:1304.1325 [gr-qc]].

\bibitem{Ilhan} 
  I.~B.~Ilhan and A.~Kovner,
  arXiv:1301.4879 [hep-th].

\bi{Woodard}
R.~P.~Woodard,
  Lect.\ Notes Phys.\  {720} (2007) 403 
  [astro-ph/0601672].

\bi{Howe-Tucker} P.S. Howe and R.W. Tucker, J. Phys. A: Math. Gen. {10}
(1977) L155.
  
\bi{PavsicSaasFee} M. Pav\v si\v c,
  Found.\ Phys.\  {35} (2005) 1617
  [hep-th/0501222].
\end{thebibliography}
\end{document}